\documentclass{ifacconf}

\usepackage{graphicx}      % include this line if your document contains figures
\usepackage{epstopdf}
\usepackage{natbib}        % required for bibliography
\usepackage{float}
\usepackage{stfloats}
\usepackage{siunitx,booktabs,array}
\usepackage{mathtools}

\usepackage{tikz}
\usepackage{atbegshi,picture}
\usetikzlibrary{positioning}
\usetikzlibrary{arrows}
\usetikzlibrary{calc}
\usepackage{amsfonts}               %
\usepackage{amsmath}                %
\usepackage{amssymb}                %
\usepackage[T1]{fontenc}

\begin{document}
\begin{frontmatter}

\title{Periodic Regulation of Linear Time-Delay Systems via Youla-Kučera Parametrization} 

\thanks[footnoteinfo]{This work was supported by the Czech Science Foundation project 24-10301S, and by the European Union under the project Robotics and advanced industrial production No. $CZ.02.01.01/00/22\_008/0004590$. The work of the first and last authors was also supported by a public grant overseen by the French National Research Agency (ANR) as part of the « Investissements d’Avenir » program, through the "ADI 2020" project funded by the IDEX Paris-Saclay, ANR-11-IDEX-0003-02.\\ \\
\textsf{\large Preprint submitted to IFAC}}

\author[CTU,PARIS]{Can Kutlu Yüksel} 
\author[CTU]{Tomas Vyhlidal}
\author[PARIS]{Silviu-Iulian Niculescu}

\address[CTU]{Department of Instrumentation and Control Engineering, Faculty of Mechanical Engineering, Czech Technical University in Prague, Technická 4, Prague 6, 16607, Czechia.}
\address[PARIS]{Universit\'{e} Paris-Saclay,
        CNRS, CentraleSup\'{e}lec, Inria, Laboratoire des Signaux et Syst\`{e}mes (L2S), 91192
        Gif-sur-Yvette, France}

\begin{abstract}                % Abstract of 50--100 words
% The paper introduces and demonstrates the effectiveness of bringing periodic regulation capibilities to already existing stabilizing controllers for time-delay system when the Youla-Kučera parametrization is considered. The scope of the paper is single-input-single-output systems. 

The paper proposes an alternative way to achieve the Internal Model Principle (IMP) in contrast to the standard way, where a model of the signal one wishes to track/reject is directly substituted into the closed-loop. The proposed alternative approach relies on an already-existing stabilizing controller, which can be further augmented with a Youla-Kučera parameter to let it implicitly admit a model of a signal without altering its stabilizing feature. Thus, with the proposed method, the standard design steps of realizing IMP are reversed. The validity and potential of the proposed approach are demonstrated by considering three different types of time-delay systems. It is shown through simulations that all considered unstable systems, despite the infinite-dimensional closed-loop, can be straightforwardly periodically regulated by augmenting their stabilizing PI controllers. Thanks to a specifically chosen structure for the Youla-Kučera parameter, the required tuning can be done by solving a set of linear equations.

% The standard way of realizing the famous Internal Model Principle(IMP)[REFER TO FRANCIS AND WONHAM] to make a closed-loop control system that can track/reject periodic signals requires the closed-loop to encapsulate the model of the signal and then its stability to be ensured.  for closed-loops involving infinite-dimensional models of both the signal and system, guaranteeing stability can be a challenging task. In this paper, we propose and demonstrate an alternative approach made possible by the Youla-Kučera parametrization. The key benefit of this parametrization is that it allows for the switching of the order of the steps of the standard IMP realization: First, stabilize the system and then augment the controller to implicitly have the signal model. Such an approach becomes especially convenient when taking into account the mass literature for the stabilization of time-delay systems. Based on this literature, periodic regulation of time-delay systems can be achieved without the challenging aspect of infinite-dimensionality. 

\end{abstract}

\begin{keyword}
periodic control \sep time-delay \sep internal model control \sep vibration control \sep disturbance rejection \sep infinite-dimensional systems.
\end{keyword}

\end{frontmatter}
%===============================================================================

\section{Introduction}
A popular way of achieving the famous Internal Model Principle (IMP) \cite{francis1976internal} to make a closed-loop control system track/reject periodic signals requires the closed-loop to encapsulate the model of the signal and then its stability to be ensured. This lets the poles of the signal model become the zeros of the closed-loop sensitivity, which in return ensures the asymptotic tracking/rejection of the targeted signal by the control system. Such a design approach is adopted by the \emph{Repetitive Control\/} method to achieve periodic regulation \cite{hara1988repetitive}. However, for closed-loop systems involving infinite-dimensional models of both the signal and the system, guaranteeing stability can be a challenging task. 

In this paper, we propose and %demonstrate 
show an alternative approach made possible by the Youla-Kučera parametrization. The key benefit of this parametrization is that it allows for the swapping of the order of the two steps in the standard IMP realization. That is, one can alternatively first stabilize the system and then augment the controller to implicitly have the signal model. Such an approach can become especially convenient when taking into account the %mass 
literature available for the stabilization of time-delay systems. In our opinion, based on this literature, periodic regulation of time-delay systems can be readily achieved without going through the challenging aspects of infinite dimensionality.

Time-delay systems can be stabilized by considering various stability analysis methods and different controller structures. See, for instance, \cite{michiels2007stability} for eigenvalue-based methods, \cite{sipahi2011stability} for graphical methods, \cite{fridman2001new} for time-domain methods, and the references therein. Among them, PID controllers can still be relevant in the stabilization of time-delay systems \cite{silva2005pid} or more recently \cite{ma2022pid,appeltans2022analysis}.

Furthermore, by letting the Youla-Kučera parameter take a particular structure involving delays, the essential regulation requirements can be posed as linear constraints and can be further optimized for additional constraints due to its optimization-friendly structure. Note that such similar methods based on parametrization were considered before for delay systems mainly in delay compensation \cite{zitek2007cascade,pekavr2007time} and in the design of repetitive controllers \cite{pipeleers2008robust,chen2013selective}.

The %presented 
proposed work can be seen as the continuation and further extension of the Internal Model Control(IMC)-based periodic regulation approach previously proposed by the authors \cite{yuksel2023harmonic,yuksel2023distributed,yuksel2025spectrum} in the recent years. It is worth mentioning that the previous works were confined to stable and minimum-phase finite dimensional systems suffering from input delays. With the proposed extension, this restriction is no longer required and, in our opinion, it represents a novelty in the open literature. 

The structure of the paper is as follows: Section 2 introduces the Youla-Kučera parametrization for single-input-single-output (SISO) linear time-invariant (LTI) systems, whether %they are time-delayed or not. 
with or without delay in the control loop. In addition, useful identities based on this parametrization are highlighted. In Section 3, a particular structure for the Youla-Kučera parameter based on time-delays is presented and is used to turn the periodic regulation requirements into linear constraints on the chosen (Youla-Kučera) structure. Section 4 demonstrates the effectiveness of the overall proposed approach via three examples, which are validated through simulations. Finally, Section 5 concludes the paper. The notations are standard and explained upon first used.

%%%
\section{Youla-Kučera Parametrization for SISO LTI Systems}
\begin{figure}[ht]
\begin{center}
\scalebox{1}{\begin{tikzpicture}
[tf/.style = {rectangle, draw=black, fill=white,  thick, minimum width=60pt, minimum height=30pt},
tf_small/.style = {rectangle, draw=black, fill=white,  thick, minimum width=30pt, minimum height=20pt},
sum/.style = {circle,draw=black,fill=white, thick, minimum size=10pt}]

% Nodes

\node[tf, label={Controller}] (C) {$C(s)$};
\node[sum] (sum_e) [left= 20pt of C] {};
\node[tf,label={ System}] (G) [right =30 pt of C] {$G(s)$};

\node[sum] (sum2) [right=15pt of G]{};
\draw(sum2.north west) --(sum2.south east);
\draw(sum2.north east) --(sum2.south west);

\node (output) [right = 20 pt of sum2] {} ;
\node (feedback) [below = 15 pt of C] {};
\node (r) [left = 15 pt of sum_e] {};

\draw[-triangle 45](sum_e.east) --(C.west) node[midway,above] () {$e$};
\draw[-triangle 45](C.east) --(G.west);
\draw[-triangle 45](G.east) --(sum2.west);
\draw[-triangle 45](sum2.east) --(output.center) node[midway] (y_label) {} node[midway,above] () {$y$};
\node[above=10pt of sum2](dd){};
\draw[-triangle 45] (dd.center) -- (sum2.north) node[midway,right]{$d$};
\draw[](y_label.center) |-(feedback.center);
\draw[-triangle 45](feedback.center) -| (sum_e.south) ;
\draw[-triangle 45] (r.center) -- (sum_e.west) node[midway,above]{$r$};

%sum_e 
\draw(sum_e.north west) --(sum_e.south east);
\draw(sum_e.north east) --(sum_e.south west);
\node[below = -2.5 pt] at (sum_e.center){-};

\end{tikzpicture}}
\end{center}
\vspace{-20pt} % Utilized to reduce the spacing between figure and its caption
\caption{Classical Feedback Control.}
\label{fig:classical}
\end{figure}
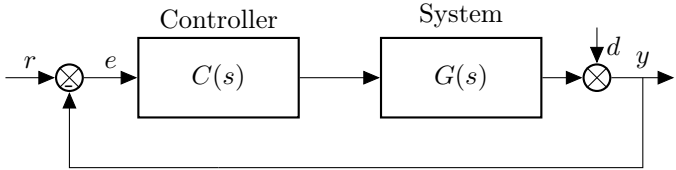

%%%
Consider the classical feedback structure shown in Fig \ref{fig:classical}, which aims to control the system $G(s)$ via controller $C(s)$. As a fundamental requirement for this system to be regulated properly, a \emph{necessary condition\/} is that the closed-loop should be stabilized by the controller.
  
Suppose that we have a controller, denoted by $C_p(s)$, which guarantees a stable closed-loop system, even if it does not fully satisfy the additional conditions imposed by the regulation requirements. Then, with respect to this controller, the Youla-Kučera parameter can capture all other stabilizing controllers, denoted generally by $C(s)$, for the system under consideration. Briefly stated, among these stabilizing controllers, our goal is to find a controller that can also satisfy our regulation requirements.

The process of deriving these controllers in a parametric form begins by representing both the system and the controller in a coprime-factorized form in terms of stable and proper transfer functions,
\begin{equation}
G(s) = \frac{N_G(s)}{D_G(s)} \quad \textrm{and} \quad C(s) = \frac{N_C(s)}{D_C(s)}
\label{eq:factorization}
\end{equation}
with $N_G(s), N_C(s), D_G(s), D_C(s)\in\mathbb{R}_{ps}(s)$. Note that $\mathbb{R}_{ps}(s)$ is the vector space of all proper and stable rational functions of the Laplace variable $s$.
Based on this factorization, the closed-loop transfer function %can be obtained as, 
writes as:
\begin{equation}
\begin{split}
        G_{cl}(s) &= \frac{G(s)C(s)}{1+G(s)C(s)} \\
        &=\frac{1}{D_G(s) D_C(s) + N_G(s) N_C(s)} N_G(s) N_C(s),
 \end{split}
\end{equation}
Notice that, by factorization, $N_G(s)$, and $N_C(s)$ are stable %systems. 
Hence, for the closed-loop to be stable (and proper), 
    \begin{equation}
        U^{-1}(s) := \frac{1}{D_G(s) D_C(s) + N_G(s) N_C(s)},
    \end{equation} must be stable (and proper). Consequently, this requirement leads to a Bézout identity,
    \begin{equation}
     D_G(s)X(s) + N_G(s)Y(s) = 1,
     \label{eq:bezout_identity}
    \end{equation} where $X(s):=D_C(s)U^{-1}(s)\in \mathbb{R}_{ps}(s)$ and $Y(s):=N_C(s)U^{-1}(s) \in \mathbb{R}_{ps}(s)$. This Bézout identity implies that if we can find such $X(s)$ and $Y(s)$, then we can use this to generate a controller that leads to a stable close-loop via
    \begin{equation}
        C(s) = \frac{N_C(s)}{D_C(s)} = \frac{Y(s)}{X(s)}.
    \end{equation} Thus, having $C_p(s)$ at hand implies that we can find $X_p(s),Y_p(s) \in \mathbb{R}_{ps}$ such that \eqref{eq:bezout_identity} is satisfied and $C_p(s) = \frac{Y_p(s)}{X_p(s)}$. Furthermore, if we happen to know such $X_p(s)$ and $Y_p(s)$, notice that for an arbitrary $Q(s)\in \mathbb{R}_{ps}$ having  $X(s) = X_p(s) - N_G(s)Q(s)$ and $Y(s) = Y_p(s) + D_G(s)Q(s) $ also satisfies the Bézout identity. Hence, one can generally express all other stabilizing controllers based on $C_p(s)$ as an affine rational function of $Q(s)$ given by,
    \begin{equation}
        C(s) = \frac{Y_p(s) + D_G(s) Q(s)}{X_p(s) - N_G(s) Q(s)}.
        \label{eq:kucera_controller}
    \end{equation}

    This arbitrary variable $Q(s)$ is referred to as the \emph{Youla-Kučera parameter} and has interesting outcomes. Among them, a prominent one is that it leads to a sensitivity transfer function that depends affinely on this parameter in contrast to $C(s)$: 

    \begin{eqnarray}
        S(s) &=& \frac{Y(s)}{D(s)}= \frac{1}{1+G(s)C(s)} \\
          &=& D_G(s) X_p(s) - D_G(s) N_G(s) Q(s).
          \label{eq:kucera_sensitivity}
    \end{eqnarray}
As a consequence, if the chosen stucture for $Q(s)$ is also affine in terms of its parameters, frequency-response optimization of the closed-loop sensitivity can be stated as a semi-definite programming \cite{iwasaki2005generalized}.

\begin{figure}[ht]
\begin{center}
\scalebox{1}{\begin{tikzpicture}[
tf/.style = {rectangle, draw=black, fill=white,  thick, minimum width=60pt, minimum height=30pt},
sum/.style = {circle,draw=black,fill=white, thick, minimum size=10pt}]

    %Nodes
\node[tf,label={above:IMC Controller}] (Q)   {$Q$};
\node[tf,label={above:System}] (G)   [right=20pt of Q]{$G(s)$} ;
\node[tf,label={above:Model}] (Gm)  [below =30pt of G.center]{$G_m(s)$};
% Sum1
\node[sum] (sum1) [left=10pt of Q]{};
\draw(sum1.north west) --(sum1.south east);
\draw(sum1.north east) --(sum1.south west);
\node[below = -2.5pt] at (sum1.center){-};

%Sum2

\node[sum] (sum2) [right=15pt of G]{};
\draw(sum2.north west) --(sum2.south east);
\draw(sum2.north east) --(sum2.south west);

\draw[-triangle 45](sum1.east) -- (Q.west);
\draw[-triangle 45](Q.east) -- (G.west) node[midway,above] {$u$} node[midway] (u) {};
\draw[-triangle 45](u.center) |-(Gm.west);
\draw[-triangle 45](G.east) -- (sum2.west);
\draw[-triangle 45](sum2.east) --++ (30pt,0pt) node[midway](output){} node[midway,above]{$y$};
\node[sum] (sum3) [below =36.5pt of output]{};
\node[below=15pt of sum3](corner){};
\node[above=10pt of sum2](dd){};
\draw[-triangle 45](output.center) -- (sum3.north);
\draw[-triangle 45](Gm.east) -- (sum3.west) node[midway,above]{$y_m$};
\draw (sum3.south) -- (corner.center);
\draw[-triangle 45] (corner.center) -| (sum1.south);
\draw[-triangle 45] (dd.center) -- (sum2.north) node[midway,right]{$d$};
\draw[triangle 45-] (sum1.west) --++(-15pt,0pt) node[midway,above]{$r$};

%Sum3

\draw(sum3.north west) --(sum3.south east);
\draw(sum3.north east) --(sum3.south west);
\node[left = -2.5pt] at (sum3.center){-};

\end{tikzpicture}}
\end{center}
\vspace{-20pt} % Utilized to reduce the spacing between figure and its caption
\caption{Internal Model Control.}
\label{fig:imc}
\end{figure}
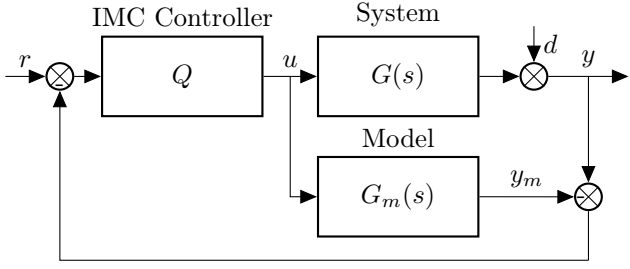

As another remark, if $G(s)$ is stable, then by letting $X_p(s) = 1$, $Y_p(s)=0$, $N_G(s) = G$ and $D_G(s) = 1$, one obtains the IMC scheme shown in Fig. \ref{fig:imc}. 
Notice that, in this case, the Youla-Kučera parameter corresponds to the IMC controller which provides another perspective on why a stable controller leads to a stable closed-loop in the IMC framework.

Before moving to the next section where we discuss how the parametrization can be utilized for the purpose of periodic regulation, notice that the original Youla-Kučera parameter $Q(s)$ stated in \eqref{eq:kucera_controller} requires the particular stabilizing controller to be factorized in a way to satisfy the Bézout identity. This constraint on the controller factorization can be relaxed by modifying the original statement of the Youla-Kučera parameter. Following the steps above, factorize the Youla-Kučera parameters in terms of an arbitrary stable and proper co-prime factorization of the controller $C_p(s)=N_p(s)D_p^{-1}(s)$ as
\begin{equation}
    Q(s) = Q_M(s) U_p^{-1}(s) \in \mathbb{R}_{ps}(s)
\end{equation} where, $U_p(s) = D_G(s) D_p(s) + N_G(s) N_p(s) \in \mathbb{R}_{ps}(s)$ and $Q_M(s) \in \mathbb{R}_{ps}(s) $ is arbitrary. $Q_M(s)$ can be seen as a modified Youla-Kučera parameter and let's all stabilizing controllers be expressed by
\begin{equation}
        C(s) = \frac{N_p(s) + D_G(s) Q_M(s)}{D_p(s) - N_G(s) Q_M(s)}.
\label{eq:kucera_controller_modified}
\end{equation}
Similar to the derivation of \eqref{eq:kucera_sensitivity}, substituting \eqref{eq:kucera_controller_modified} into sensitivity expression for the classical feedback yields an affine function of the modified Youla-Kučera parameter $Q_M(s)$ as
\begin{equation}
    S(s) = U_p^{-1}(s) D_G(s) (D_p(s) - N_G(s) Q_M(s)).
\label{eq:kucera_sensitivity_modified}
\end{equation}

Based on Equation \eqref{eq:kucera_sensitivity_modified}, we can make the following two categorization of the sensitivity zeros:
\begin{enumerate}
    \item Open-loop poles of the system become sensitivity zeros due to $D_G(s)$.
    \item Zeros of $D_p(s)-N_G(s)Q_M(s)$ become zeros of the sensitivity. 
\end{enumerate}

Thus, based on the second category, we can place zeros through the Youla-Kučera parameter without altering the stability of the original closed-loop system. 

\section{Weighted Lumped-Delays as Youla-Kučera Parameter}

Essentially, what IMP states is that the sensitivity transfer function should attain zeros at the harmonic frequencies that constitute the targeted signal. That is, if the signal desired to be asymptotically tracked/rejected has a finite Fourier series expansion, as in
\begin{equation}
    %v(t) = \sum_{i=-k}^{k} a_i \euler^{-\mathrm{j} \omega\frac{2\pi i}{T}}.
    v(t) =\frac{c_0}{2}+\sum_{l=1}^{M_d} c_l\cos\left(\frac{2\pi l}{T}t-\varphi_l\right),
    \label{eq:Fourier}
\end{equation}
then the sensitivity should satisfy
\begin{equation}
    S(0)= S(\mathrm{j} \omega_l) = 0,
    \label{eq:sensitivity_cond}
\end{equation} where $\omega_l = \frac{2\pi l}{T}$ and $l = 1...M_d$.

Based on Equation \eqref{eq:kucera_sensitivity_modified}, this condition on sensitivity can be projected on the modified parameter $Q_M(s)$ as
\begin{equation}
            Q_M(j\omega_l) = D_p(j\omega_l)N_G^{-1}(j\omega_l).
            \label{eq:yk_zero_condition}
\end{equation}

Condition \eqref{eq:yk_zero_condition} can be turned into a linear condition on the sub-parameters of $Q_M(s)$ if we let
\begin{equation}
    Q_M(s) =\sum_{k=0}^N a_k e^{-s\tau_k} \in \mathbb{R}_{ps}(s),
    \label{eq:yk_structure}
\end{equation} by which $Q_M(s)$ is characterized by parameters $a_k$, which scale $N$ past signal values, each separated by a time interval of $\vartheta$, i.e. $\tau_k=k \vartheta$, resulting in a delay effect that spans for a total time of $T_D =\vartheta N$. With this decision, condition \eqref{eq:yk_zero_condition} can be achieved by finding $a_k$ for a properly chosen set of delays by solving a set of linear equations,
\begin{equation}
    Ax=B,
    \label{eq:yk_linear_cond}
\end{equation}
where $x\in\mathbb{R}^{N+1}$, given as $x=[a_0, a_1,..., a_{N}]^\mathrm{T}$, and $A\in\mathbb{R}^{2M_d+1\times N+1}, B\in\mathbb{R}^{2M_d+1}$,
\begin{equation*}
A=\left[
    \begin{array}{cccc}
     1 & 1 & \cdots& 1 \\
      1   & \cos(\omega_1\vartheta) & \cdots & \cos(\omega_1 N\vartheta) \\
     \vdots & \vdots & \cdots& \vdots \\
    1   & \cos(\omega_{M_d} \vartheta) &\cdots & \cos(\omega_{M_d} N\vartheta) \\
    0   & \sin(\omega_1 \vartheta) &\cdots & \sin(\omega_1  N\vartheta) \\
     \vdots & \vdots & \cdots& \vdots\\
    0   & \sin(\omega_{M_d}  \vartheta) & \cdots & \sin(\omega_{M_d}  N\vartheta) \\
    \end{array}
    \right], \,
B=\left[
\begin{array}{c}
R_0 \\ R_1 \\ \vdots \\ R_{Md} \\ I_1 \\ \vdots \\ I_{Md}
\end{array}
 \right].
    \end{equation*}
with $R_0=D_p(0)N_G^{-1}(0)$ and
\begin{equation*}
  R_l=\Re\left( D_p(j\omega_l)N_G^{-1}(j\omega_l)  \right), I_l=-\Im\left( D_p(j\omega_l)N_G^{-1}(j\omega_l)\right) 
\end{equation*}
for $l=1,2,\dotsc,M_d$. Recall that for the linear set of equations \eqref{eq:yk_linear_cond} to have a solution, matrix $A$ must have full row-rank with $N\geq2M_d$. Provided this, a solution can be found using either the Moore-Penrose inverse as in
\begin{equation}
    x = (A^TA)^{-1}A^TB
    \label{eq:pinv}
\end{equation}
or through an optimizer if the solution has to satisfy further constraints. 

\begin{figure}[ht]
\begin{center}
\scalebox{1}{\begin{tikzpicture}[tf/.style = {rectangle, draw=black, fill=white,  thick, minimum width=60pt, minimum height=30pt},
tf_small/.style = {rectangle, draw=black, fill=white,  thick, minimum width=30pt, minimum height=20pt},
sum/.style = {circle,draw=black,fill=white, thick, minimum size=10pt}]

\node[sum] (sum_1) {};

\node[sum] (sum_2) [right = 10 pt of sum_1] {};
\node[tf_small] (x_p) [right = 10 pt of sum_2] {$D^{-1}_p$}; 
\draw[-triangle 45](sum_1.east) -- (sum_2.west);

\draw[-triangle 45](sum_2.east) -- (x_p.west);

\node[tf_small] (y_p) [right = 20 pt of x_p] {$N_p$};
\node[sum] (sum_3) [right = 10 pt of y_p] {};
\draw[-triangle 45](y_p.east) -- (sum_3.west);
\node (x_c) [right = 5 pt of x_p] {};
\draw[-triangle 45](x_p.east) -- (y_p.west);
\node[tf_small] (q) [below = 20 pt of x_c] {$Q_M$} ; 
\draw[-triangle 45](x_c.center) -- (q.north);
\node (y_c) [below = 10 pt of q] {};
\draw[](q.south) -- (y_c.center);

\node[tf_small] (n_g) [below= 40 pt of x_p] {$N_G$};
\node[tf_small] (d_g) [below= 40 pt of y_p] {$D_G$};
\draw[-triangle 45] (y_c.center) |- (n_g.east);
\draw[-triangle 45] (y_c.center) |- (d_g.west);

\draw[-triangle 45] (n_g.west) -| (sum_2.south);
\draw[-triangle 45] (d_g.east) -| (sum_3.south);

\node[tf_small] (g) [right = 10 pt of sum_3] {$G(s)$};
\draw[-triangle 45](sum_3.east) -- (g.west);

\node[sum] (sum_4) [right = 10 pt of g] {};

\node[above=10pt of sum_4](dd){};
\draw[-triangle 45] (dd.center) -- (sum_4.north) node[midway,right]{$d$};
\draw[-triangle 45] (g.east) -- (sum_4.west);

\node (output) [right = 20 pt of sum_4] {} ;
\draw[-triangle 45](sum_4.east) --(output.center) node[midway] (y_label) {} node[midway,above] () {$y$};

\node (feedback) [below = 10 pt of d_g] {};
\draw[](y_label.center) |-(feedback.center);
\draw[-triangle 45](feedback.center) -| (sum_1.south) ;

\node (r) [left = 15 pt of sum_1] {};
\draw[-triangle 45] (r.center) -- (sum_1.west) node[midway,above]{$r$};

% \node (m) [below = 20 pt of x_p] {};
% \node[tf_small] (N) [right= 2pt of m] {$N_G(s)$};
% \draw[-triangle 45] (x_c.center) |- (N.east);
% \node[tf_small] (W_back) [left = 2 pt of m] {$Q(s)$};
% \draw[-triangle 45] (N.west) |- (W_back.east);
% \draw[-triangle 45] (W_back.west) -| (sum_3.south);

% \node[tf] (y_p) [right = 100 pt of sum_3] {$Y_p(s)$};
% \draw[-triangle 45] (sum_2.east) -- (G.west) node[midway] (x_c) {};

% \draw[-triangle 45] (sum_1.east) -- (y_p.west) node[midway] (y_c) {};
% \node (n) [below = 20 pt of y_p] {};
% \node[tf_small] (W_forward) [left = 2 pt of n] {$Q(s)$};
% \draw[-triangle 45] (y_c.center) |- (W_forward.west);
% \node[tf_small] (D) [right= 2pt of n] {$D_G(s)$};
% \draw[-triangle 45] (W_forward.east) -- (D.west);
% \node[sum] (sum_2) [right = 15pt of y_p] {};
% \draw[-triangle 45] (y_p.east)--(sum_2.west);
% \draw[-triangle 45] (D.east) -| (sum_2.south);
% \node[tf] (G) [right = 50 pt of sum_2.west] {$G(s)$};
% \draw[-triangle 45] (sum_2.east) -- (G.west) node[midway] (x_c) {};

% \node (r) [left = 15 pt of sum_1] {};

% \node (y) [right = 20 pt of G] {};
% \draw[-triangle 45] (G.east) -- (y.center) node[midway] (y_mid) {};
% \node (back) [below = 50pt of G] {};
% \draw[] (y_mid.center) |- (back.center) ;
% \draw[-triangle 45] (back.center) -| (sum_1.south) ;
% \draw[-triangle 45] (r.center) -- (sum_1.west);

%sum_1
\draw(sum_1.north west) --(sum_1.south east);
\draw(sum_1.north east) --(sum_1.south west);
\node[below = -2.5 pt] at (sum_1.center){-};

%sum_2
\draw(sum_2.north west) --(sum_2.south east);
\draw(sum_2.north east) --(sum_2.south west);
%\node[below = -1 pt] at (sum_2.center){-};

%sum_3
\draw(sum_3.north west) --(sum_3.south east);
\draw(sum_3.north east) --(sum_3.south west);
%\node[below = -1 pt] at (sum_3.center){-};

%sum_4
\draw(sum_4.north west) --(sum_4.south east);
\draw(sum_4.north east) --(sum_4.south west);

\draw[thick,dotted] ($(sum_2)+(-13pt, -75pt)$) rectangle ($(sum_3)+(8pt, +25pt)$) node[below= 10pt,left] {$C(s)$};

\end{tikzpicture}}
\end{center}
\vspace{-20pt} % Utilized to reduce the spacing between figure and its caption
\caption{Classical Feedback Control in Youla-Kučera form.}
\label{fig:kucera}
\end{figure}
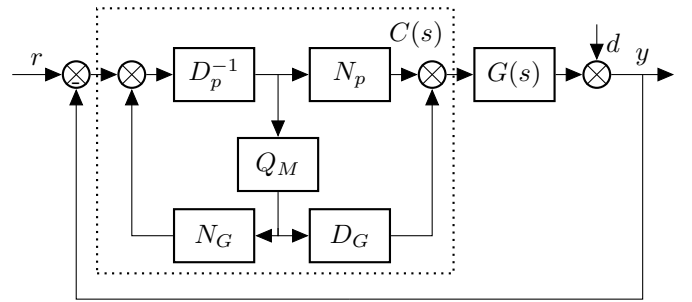

Before moving to the next section, where examples are provided, let us highlight the alternative steps based on this parametrization to gain a controller capable of periodically regulating a system:
\begin{enumerate}
    \item Find a stabilizing controller $C_p(s)$ for the system $G(s)$.
    \item Factorize both the stabilizing controller and the system as a coprime ratio of stable and proper systems.
    \item Based on this factorization, find the gains for $Q_M(s)$ structured as in \eqref{eq:yk_structure} so that it satisfies condition \eqref{eq:yk_zero_condition}.
   \item Augment the stabilizing controller as depicted in Fig. \ref{fig:kucera}.
\end{enumerate}

\section{Demonstrative Examples}

\begin{figure}[h!]
    \centering
    \includegraphics[width = 0.95\columnwidth]{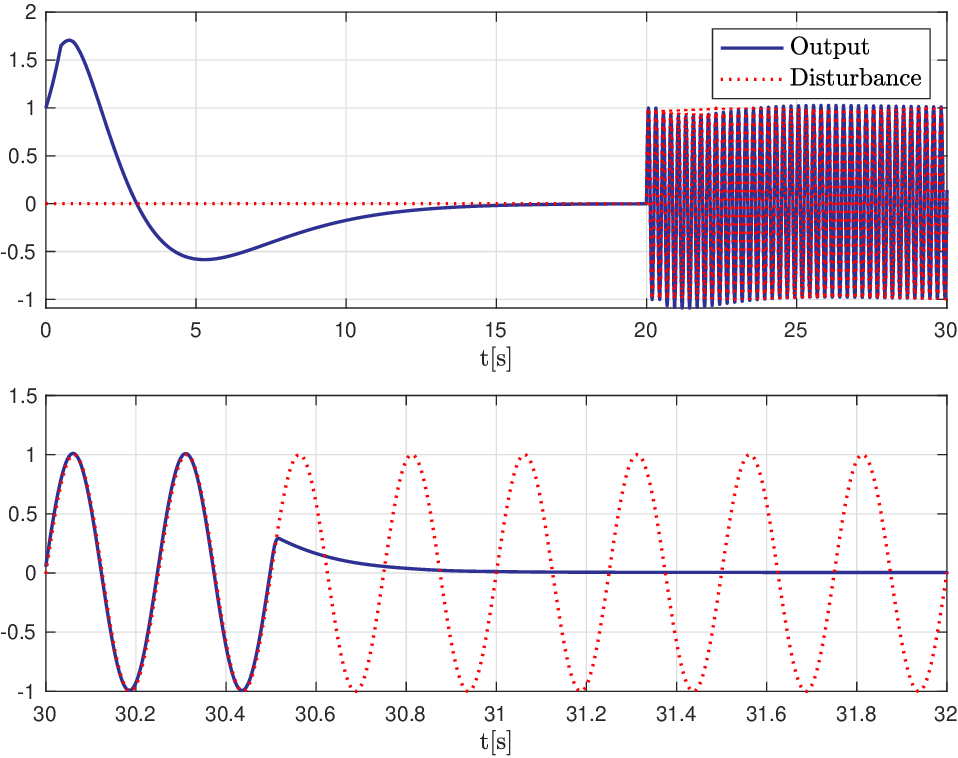}
    \caption{\textbf{(Top)} Stabilization and Rejection Performance of the PI Controller for the Unstable System with Input Delay in \eqref{eq:G_unstable} \textbf{(Below)} Improved Rejection Performance of the Augmented PI Controller with Youla-Kučera parameter. }
    \label{fig:first_response}
\end{figure}

In this section, the proposed structure for the parameter in Equation \eqref{eq:yk_structure} is utilized and demonstrated to show how one can make time-delay systems regulated periodically. We begin our demonstration with a rather simple-looking but challenging problem:

\textbf{Example 1:} Consider the unstable system of the form
\begin{equation}
         G(s) = \frac{Y(s)}{U(s)} = \frac{1}{s-a}e^{-s\tau},
    \label{eq:G_unstable}
\end{equation} where $a>0$ and $\tau>0$. As stated in \cite{ma2022pid}, PID controllers can stabilize such systems if the controller gains $K_P$, $K_I$, and $K_D$ are chosen to satisfy a certain condition. For a numerical demonstration, suppose the considered system has $a = 1$ and $\tau=0.5$. A suitable PID controller for this system can then be found, for instance, as $K_P=1.27$,  $K_I=0.0536$, and $K_D = 0$.  As for the Youla-Kučera parametrization, the PID controller $C_p(s)$ can be arbitrarily expressed as a ratio of two coprime stable transfer functions 
 \begin{equation}
 \begin{cases}
 N_p(s) = \frac{K_P s + K_I}{s+1},\\
 D_p(s) = \frac{s}{s+1}.
      \label{eq:pi_coprime}
 \end{cases}
 \end{equation}
 The same can be done for the time-delay system in \eqref{eq:G_unstable} as
 \begin{equation}
 \begin{cases}
     N_G(s) = \frac{\mu}{s^2 + 2 \sqrt{\mu} s + \mu} e^{-s\tau}, \\
     D_G(s) = \frac{(s-1)\mu}{s^2 + 2 \sqrt{\mu} s + \mu},
 \end{cases}
 \end{equation}
 where $\mu>0$ is a simple parametrization of stable second-order polynomials with repeated roots. Notice that the system delay is associated with $N_G(s)$ rather than $D_G(s)$ to keep the coprime factorization stable and causal. Under these settings, the particular Youla-Kučera parameter in (\ref{eq:yk_structure}) can be tuned to satisfy the periodic regulation condition given by (\ref{eq:yk_zero_condition}):
 \begin{equation}
    \sum_{k=0}^N a_k e^{-j\omega_l\tau_k} = \frac{j\omega_l}{j\omega_l+1} \frac {{j\omega_l}^2 + 2 \sqrt{\mu} j\omega_l + \mu}{\mu} e^{j\omega_l\tau}
    \label{}
\end{equation} for $l=0,...,M_d$. Thus, for the sake of demonstration, if we target a single harmonic signal with frequency $f = 4 [Hz]$, we can let $Q_M(s)$ in \eqref{eq:yk_structure} be characterized by $N = 2$, $\tau_k=0.01k$ and $\mu=100$. Based on \eqref{eq:yk_zero_condition}, these settings pose three linear equations with three unknowns corresponding to the coefficients $a_k$, which can be practically solved with a hand calculator. The following regulation performance of the particular unstable time-delay system obtained via straightforward calculation of the Youla-Kučera parameter is presented in Fig. \ref{fig:first_response}.

The original control system is initiated from a non-zero initial condition, proving the stabilizing effect of the PI controller and later on gets exposed to the single harmonic disturbance at $t=20[s]$, demonstrating the incapability of the PI controller to suppress the disturbance. At $t=30[s]$, the augmented part of the PI controller is activated, letting a perfect suppression of the disturbance after $0.5$ seconds of delay imposed by $N_G(s)$.  

\begin{figure}[h!]
    \centering
    \includegraphics[width = 0.95\columnwidth]{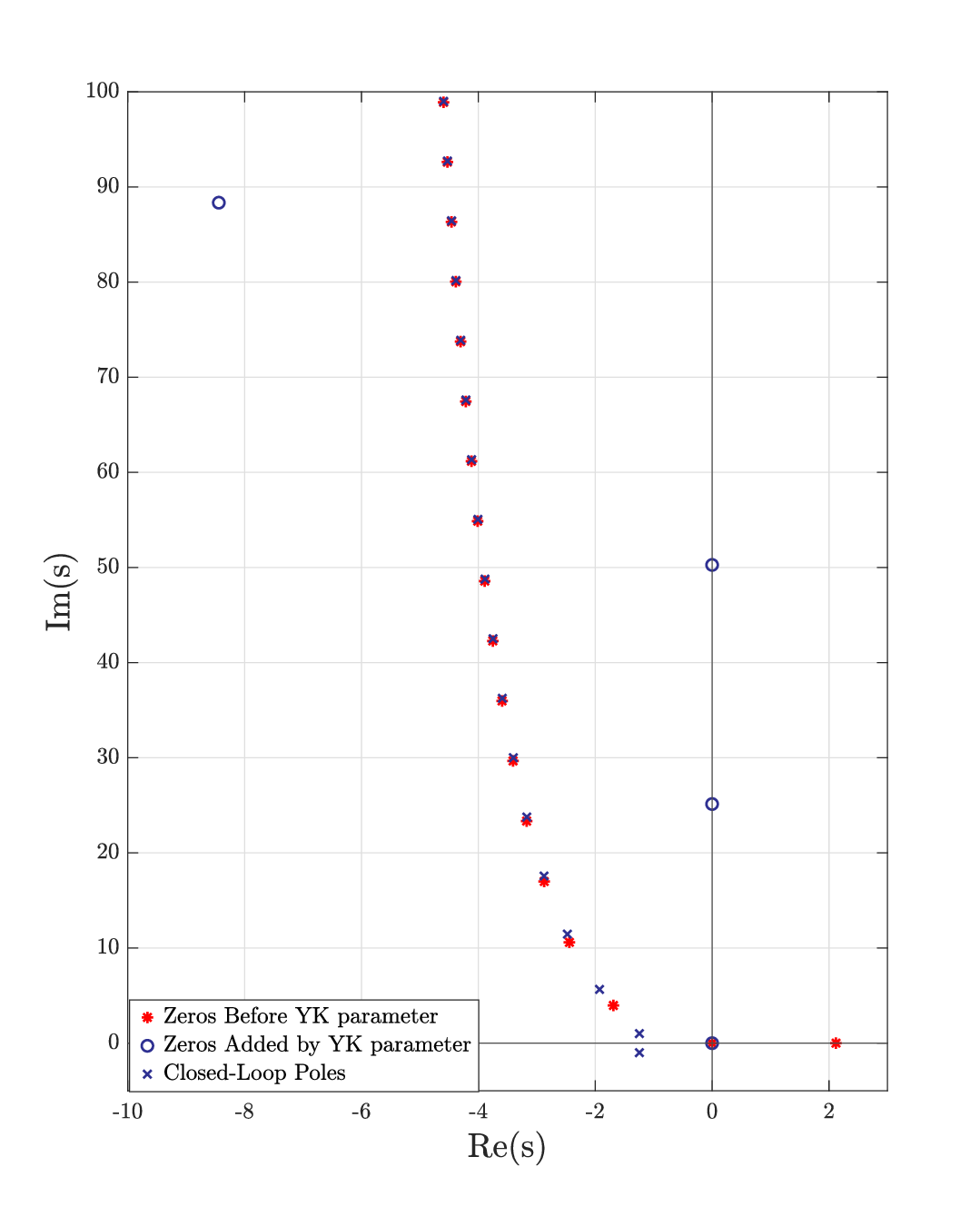}
    \vspace{-24pt}
    \caption{Pole and Zero Spectrum of the Sensitivity for the Retarded Closed-loop System from Example 2.}
    \label{fig:retarded_spectrum}
\end{figure}

\textbf{Example 2:} Consider the retarded time-delay system \begin{equation}
    G(s) = \frac{Y(s)}{U(s)} = \frac{1}{s-2-e^{-s}},
    \label{eq:retarded_sys}
\end{equation} which happens to be unstable. We can validate that the unstable retarded time-delay system can be stabilized by a PI controller that has the gains $K_P= 10$ and $K_I= 10$ by checking the resulting spectrum of the closed-loop transfer function,
\begin{equation}
    G_{cl}(s) = \frac{10s+10}{s^2+8s+10-se^{-s}},
\end{equation}
generated via the \texttt{QPmR} algorithm \cite{vyhlidal2014qpmr} and shown in Fig. \ref{fig:retarded_spectrum}. Note that the open-loop poles correspond to the zeros of the sensitivity without the Youla-Kučera parameter and the unstable behavior is caused by the single pole on the right-hand side of the complex domain.

\begin{figure}[h!]
    \centering
    \includegraphics[width = 0.95\columnwidth]{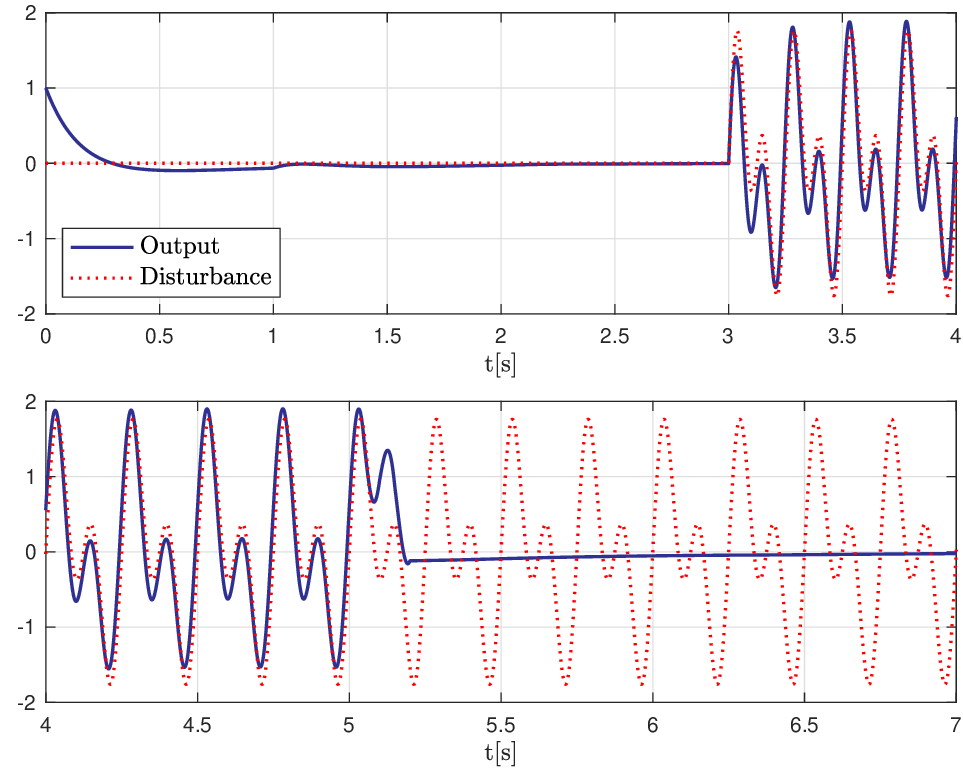}
    \caption{\textbf{(Top)} Stabilization and Rejection Performance of the PI Controller for the Unstable Retarded System given by \eqref{eq:retarded_sys} \textbf{(Below)} Improved Rejection Performance of the Augmented PI Controller with Youla-Kučera parameter. }
    \label{fig:retarded_response}
\end{figure}

Similar to the previous example, we can augment this stabilizing PI controller by the Youla-Kučera parametrization of the form \eqref{eq:yk_structure} to bring a multi-harmonic periodic regulation feature that aims for $f =4[Hz]$ with $M_D = 2$. In this case, we can let the coprime factorization of the controller be exactly as in \eqref{eq:pi_coprime}, and the coprime factorization of the system \eqref{eq:retarded_sys} can be carried out as
\begin{equation}
    \begin{cases}
    N_G(s) = \frac{1}{s+1}, \\
    D_G(s) = \frac{s-2-e^{-s}}{s+1}.
    \end{cases}
    \label{eq:retarded_coprime}
\end{equation}
Based on these factorizations and the chosen structure for the Youla-Kučera parameter in \eqref{eq:yk_structure}, we can find the coefficients that meet the periodic regulation requirements by solving the set of linear equations given by 
\begin{equation}
    \sum_{k=0}^N a_k e^{-j\omega_l\tau_k} = \frac{j\omega_l}{j\omega_l+1} \frac{j\omega_l+1}{1}= j\omega_l.
\end{equation} Setting $N=4$ with $\tau_k=0.05k$ leads to a square matrix $A$ with full-rank. Thus, the unknown coefficients can be found as $a_0=0$, $a_1=-21.3792$, $a_2=13.2131$, $a_3 = -13.2131 $ and $a_4= 21.3792 $. The resulting spectrum in Fig. \ref{fig:retarded_spectrum} clearly shows the placed harmonic zeros without affecting the stability of the closed loop. The achieved regulation performance of the unstable retarded system against a two-harmonic periodic disturbance is provided in Fig. \ref{fig:retarded_response}. The stabilizing effect of the PI controller is demonstrated by initiating the system from a non-zero initial condition. The two-harmonic periodic disturbance is introduced at $t=3[s]$, which the PI controller alone is incapable of suppressing. At $t=5[s]$, the augmented part for the controller is activated, which introduces the required dynamics for a successful suppression. 


\begin{figure}[h!]
    \centering
    \includegraphics[width = 0.95\columnwidth]{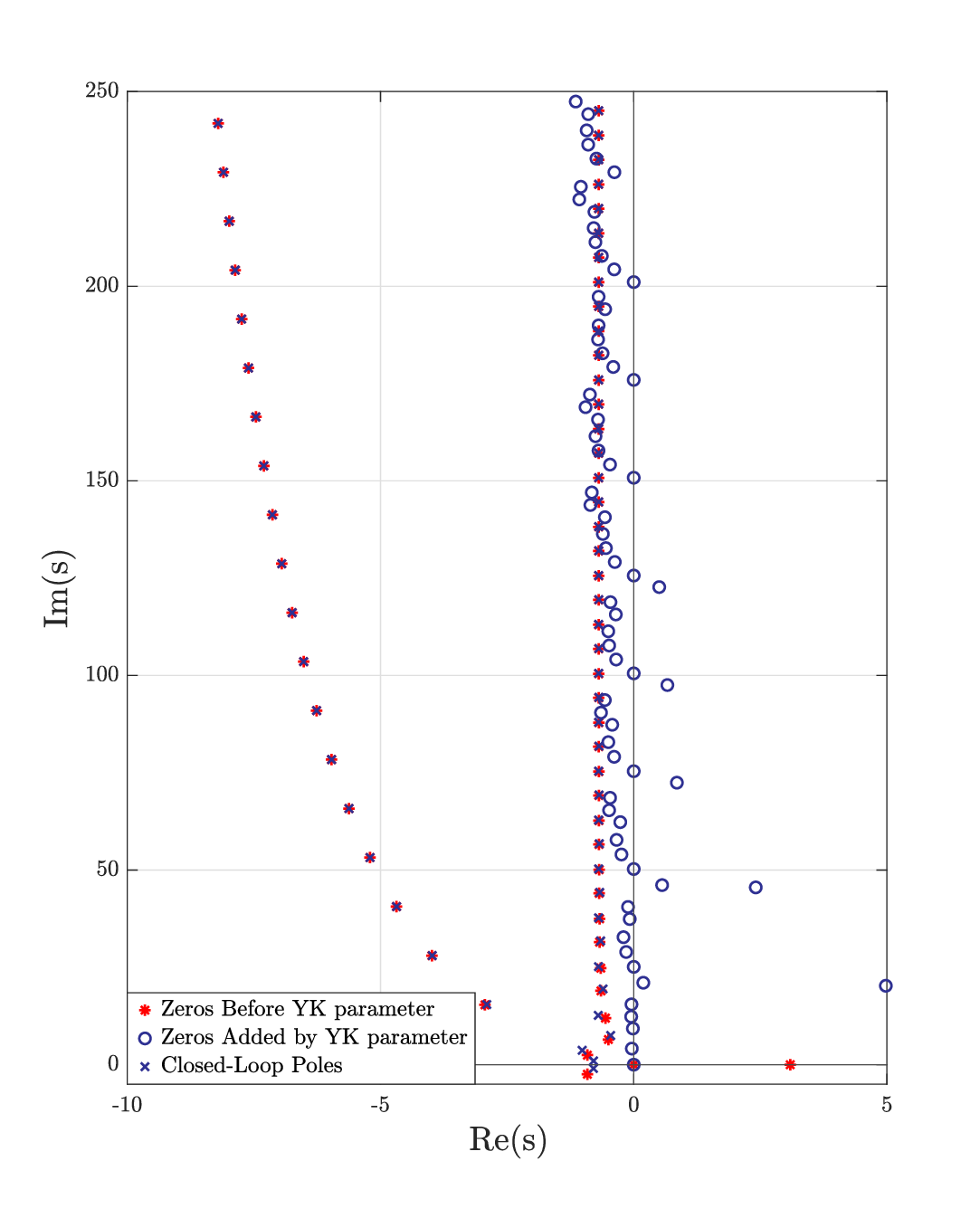}
    \vspace{-24pt}
    \caption{Pole and Zero Spectrum of the Sensitivity for the Neutral Closed-loop System from Example 3.}
    \label{fig:neutral_spectrum}
\end{figure}

\textbf{Example 3:} Consider the neutral system 
\begin{equation}
    G(s) = \frac{Y(s)}{U(s)} = \frac{1}{s(1-0.5e^{-s})-2e^{-1.5s}-3},
    \label{eq:neutral_sys}
\end{equation} which is unstable but has an exponentially stable associated difference equation. As a consequence, the neutral time-delay system is unstable due to finitely many unstable poles, as can be clearly seen in Fig. \ref{fig:neutral_spectrum}, \cite{michiels2005eigenvalue}. Similar to Example 2, this neutral system can also be stabilized by a PI controller with gains $K_P= 10$ and $K_I= 10$. If we apply the same factorization of this PI controller as in \eqref{eq:pi_coprime} and furthermore let the neutral system be decomposed as 
\begin{equation}
    \begin{cases}
    N_G(s) = \frac{1}{s+1}, \\
    D_G(s) = \frac{s(1-0.5e^{-s})-2e^{-1.5s}-3}{s+1},
    \end{cases}
    \label{eq:retarded_coprime}
\end{equation}
then notice that based on \eqref{eq:yk_zero_condition}, the same Youla-Kučera parameter evaluated in Example 2 can be directly utilized in Example 3 to make the system track/reject a signal with $f =4[Hz]$ with $M_D = 2$. Thus, under certain conditions, one particular Youla-Kučera parameter can make several different systems attain periodic regulation features. 

\begin{figure}[h!]
    \centering
    \includegraphics[width = 0.95\columnwidth]{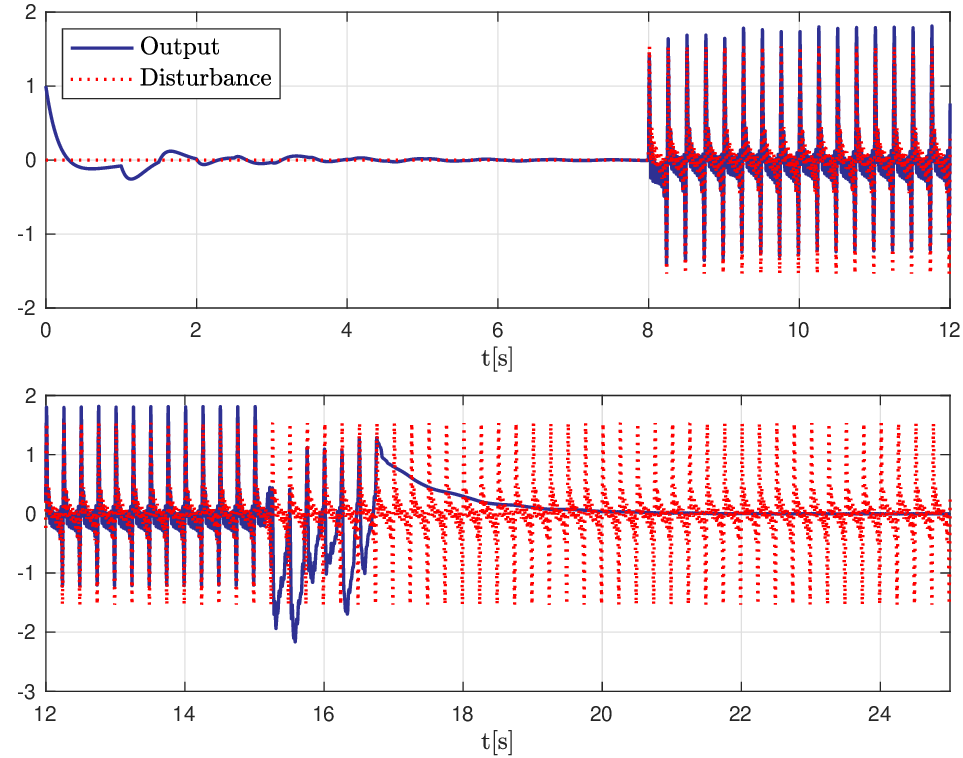}
    \vspace{-10pt}
    \caption{\textbf{(Top)} Stabilization and Rejection Performance of the PI Controller for the Unstable Neutral given by \eqref{eq:neutral_sys} \textbf{(Below)} Improved Rejection Performance of the Augmented PI Controller with Youla-Kučera parameter. }
    \label{fig:neutral_response}
\end{figure}

Nevertheless, to demonstrate the extent of multi-frequency regulation, let's consider $f =4[Hz]$ with $M_D = 8$. Setting $N=25$ with $\tau_k=0.08k$ results in a matrix $A$, which is wide with full-row rank. Thus, there exist infinitely many options for the decision of the coefficients $a_k$. One particular solution, obtained by using the Moore-Penrose inverse \eqref{eq:pinv}, yields the spectral distribution shown in Fig. \ref{fig:neutral_spectrum}. Notice that the sensitivity spectrum of the neutral closed-loop system attains infinitely many new zeros imposed by the resulting Youla-Kučera parameterization. It should be noted that the distribution of these added zeros depends on $N$ and $\tau_k$. Nevertheless, a closer look at this distribution reveals that all targeted harmonic zeros are placed at their respective positions on the imaginary axis.
However, it should be noted that the additional zeros introduced by the Youla-Kučera parameter can alter other important aspects, such as the robustness and transient performance of the closed-loop, and, therefore, still needs careful handling. Such further requirements or constraints on these properties can be further explored in an optimization-based framework. The resulting regulation performance of the augmented PI controller obtained via \eqref{eq:pinv} is demonstrated in Fig. \ref{fig:neutral_response}.
As before, the stabilization feature of the PI controller is demonstrated by initializing the closed-loop from a non-zero point. The aggressive disturbance, comprising eight harmonics, starts acting at $t=8[s]$, revealing once again the incapability of the PI controller for suppression. Nevertheless, with the augmented part activated at $t=15[s]$, the disturbance is eventually removed. The transient performance of the regulation suggests further requirements for tuning the coefficients in order to achieve a smoother transition to steady-state.

\section{Conclusion}

Thanks to the Youla-Kučera parametrization, periodic regulation of time-delay systems can be straightforwardly achieved if there exists a stabilizing controller for the considered system. Furthermore, by imposing a certain structure on this parameter, which is based on utilizing time-delays, the tuning of the Youla-Kučera parameter can be carried out by solving a set of linear equations. Three simulation-based case studies back these observations. Nevertheless, despite the augmented controller has an inherent robustness against small system/model mismatches thanks to encapsulating a model of the system, the additional zeros introduced through Youla-Kučera parametrization suggest further tuning criteria to ensure the robustness and transient response of the overall closed-loop remain feasible. Improving the tuning of the parametrization together with the generalization of the parametrization for periodic regulation of MIMO time-delay systems and experimental validation remains to be further explored as future work.

\bibliography{ifacconf} 
\end{document}